# Starbugs: all-singing, all-dancing fibre positioning robots

James Gilbert[*], Michael Goodwin, Jeroen Heijmans, Rolf Muller, Stan Miziarski, Jurek Brzeski,
Lew Waller, Will Saunders, Alex Bennet, Julia Tims

Australian Astronomical Observatory, PO Box 296, Epping NSW 1710, Australia

**ABSTRACT**

Starbugs are miniature piezoelectric 'walking' robots with the ability to simultaneously position many optical fibres across a telescope's focal plane. Their simple design incorporates two piezoceramic tubes to form a pair of concentric 'legs' capable of taking individual steps of a few microns, yet with the capacity to move a payload several millimetres per second. The Australian Astronomical Observatory has developed this technology to enable fast and accurate field reconfigurations without the inherent limitations of more traditional positioning techniques, such as the 'pick and place' robotic arm. We report on our recent successes in demonstrating Starbug technology, driven principally by R&D efforts for the planned MANIFEST (many instrument fibre-system) facility for the Giant Magellan Telescope. Significant performance gains have resulted from improvements to the Starbug system, including i) the use of a vacuum to attach Starbugs to the underside of a transparent field plate, ii) optimisation of the control electronics, iii) a simplified mechanical design with high sensitivity piezo actuators, and iv) the construction of a dedicated laboratory 'test rig'. A method of reliably rotating Starbugs in steps of several arcminutes has also been devised, which integrates with the pre-existing $x$-$y$ movement directions and offers greater flexibility while positioning. We present measured performance data from a prototype system of 10 Starbugs under full (closed-loop) control, at field plate angles of 0-90 degrees.

## 1. INTRODUCTION

The next class of Extremely Large Telescopes (ELTs) will begin a new era of unprecedented science in ground-based astronomy, but with this comes a number of new challenges for telescope systems and instrumentation. One such challenge is the efficient reconfiguration of focal planes for fibre-fed multi-object astronomy, where many optical fibres must be precisely placed on specific objects within the telescope's field of view. Existing fibre positioning technologies exhibit major limitations when applied to large focal planes such as those of ELTs. Current 'pick and place' robotic arm systems, such as 2dF on the Anglo-Australian Telescope[1] (Figure 1-a) and OzPoz on the Very Large Telescope[2], position fibres sequentially, meaning that the time taken for a field reconfiguration increases linearly with the number of fibres present. Total configuration times therefore become unacceptably long for large telescopes with a high multiplex of objects. The Australian Astronomical Observatory (AAO) has developed Starbug technology to overcome this limitation and provide a multitude of additional advantages over existing positioning systems.

Starbugs are miniature piezoelectric robots that can quickly and accurately position many optical 'payloads' (e.g. fibres) simultaneously. They were first described in 2004[3], and later in 2006[4][5] and 2010[6]. An individual Starbug comprises two piezoceramic tube actuators, joined at one end to form a pair of concentric 'legs' that can be electrically driven to produce a micro-stepping motion in the $\pm x$ and $\pm y$ directions (i.e. forwards, backwards, left, right) (Figure 1-b). Discrete step sizes of only a few microns mean that Starbugs are precise, yet they can move several millimetres per second when stepping at high frequencies. Each Starbug carries an individual payload such as an optical fibre, fibre bundle or microlens assembly; these are mounted in the Starbug's central aperture. The Starbugs operate in an 'inverted hanging' format (Figure 1-c) in order to eliminate the need for fibre retractors. This arrangement employs a transparent glass field plate, through which starlight passes before entering the fibre feed optics in the centre of each Starbug. Previous Starbug designs used a pair of ring-shaped magnets to clamp them to the field plate[6] (Figure 1-d), but this has since been replaced with a novel vacuum system, presented in this paper. Starbug technology is patent pending.

*james.gilbert@aao.gov.au; phone +61 2 9372 4833; www.aao.gov.au

Simultaneous positioning of all Starbugs means that an entire focal plane can be reconfigured in a matter of minutes, regardless of how many Starbugs there are. By contrast, the 2dF pick and place robot takes of order one hour to sequentially reposition 400 fibres and requires a field plate exchange system with duplicate fibre bundles in order to minimise observing downtime[1]. Further merits of a Starbug system include the possibility of reconfiguring groups of fibres while other groups are observing, and micro-tracking objects during observations.

Recent R&D efforts for Starbugs have been driven by MANIFEST (many instrument fibre-system) for the Giant Magellan Telescope (GMT). MANIFEST[7] is a planned fibre positioning facility for feeding multiple instruments from the Gregorian focus of this 24.5 m ELT. Starbugs are particularly suited to MANIFEST's challenging requirements and are the AAO's chosen positioning technology for this instrument.

In the following sections: we describe the design of our latest prototype Starbugs and their control system; we present results of extensive laboratory testing, including measured Starbug performance in terms of speed, resolution and payload mass; we report on our success in demonstrating fast parallel field reconfigurations with multiple Starbugs on a vertical surface; and we address reliability and maintenance factors such as Starbug lifespan, field plate endurance and vacuum system robustness. The Starbugs system for the MANIFEST instrument is also discussed in more detail.

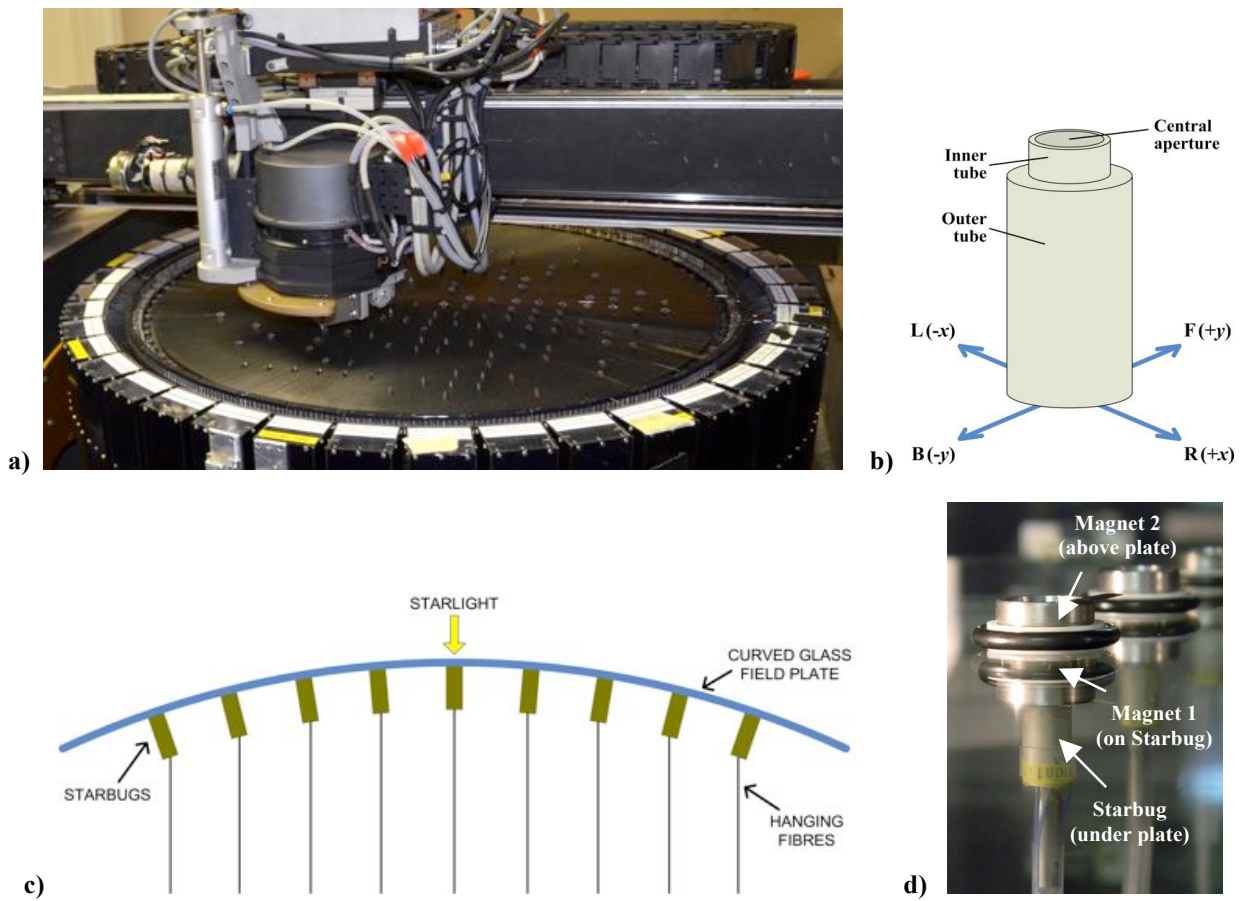

Figure 1: a) The 2dF pick and place robot has a mechanical gripper which sequentially positions up to 400 fibres on the focal plane of the Anglo-Australian Telescope; b) Starbugs nominally move in orthogonal *x-y* directions (forwards , backwards, left, right) and have a clear central aperture for inserting instrument optics; c) Starbugs are intended to operate in an inverted format, positioning underneath a (flat or curved) transparent field plate and are held to the plate with a vacuum (drawing not to scale); d) the original Starbug format used a pair of ring shaped magnets to 'clamp' them to the field plate.

## 2. THEORY OF OPERATION

The operation of Starbugs is based on the deflection modes of piezo tube actuators, with a vacuum used to attach the devices to the field plate, as described below.

### 2.1 The discrete stepping Starbug

Starbugs use a pair of piezoceramic tube actuators to produce a micro-stepping motion. Their construction is remarkably simple: one piezo tube is mounted within another piezo tube using a solid ring at one end. At the opposite end of this assembly, the ends of the tubes are polished flat. The application of an electrical potential across the walls of these tubes will cause deformations which can be exploited to produce a stepping motion (described in detail in 2010[6]). An animation of the stepping sequence can be viewed in Video 1. Previous Starbugs used a trapezoidal waveform shape to produce their stepping motion, but sinusoidal waveforms have since been adopted for better electrical efficiency. Four of these waveforms are required to drive one Starbug (two per tube), and must have amplitudes of ~200 V in order to displace the piezo tubes by sufficient amounts. Starbugs can walk on flat or curved surfaces, but require low surface roughness due to their actuator movements being so small (order 10 µm). The actual displacement of a piezo tube is dependent on its dimensions, such that a long tube with small diameter (i.e. a tall and narrow Starbug) will yield a larger (maximum) step size and therefore enable faster positioning. A fundamental trade-off in a Starbug's design is in balancing the maximum step size with mechanical stability, where shorter and wider Starbugs are more stable.

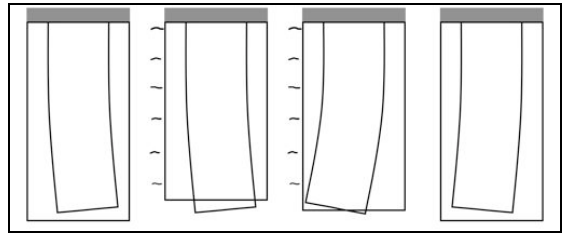

Video 1: Animation of a Starbug's discrete stepping motion (exaggerated);
http://proceedings.spiedigitallibrary.org/data/Conferences/SPIEP/68021/PSISDG_8450_84501A_ds001.mov

The direction in which a Starbug moves is controlled by applying waveforms to different electrodes on its inner piezo tube. This tube has four equally spaced electrodes around its outer wall, meaning that steps can be made in four orthogonal directions ($\pm x$ and $\pm y$). Alternatively, the tube could be divided into three electrodes to give three movement directions; this would reduce electrical complexity but at a cost of flexibility while positioning. A simple method of rotating a Starbug about its centre has recently been developed based on the same principles as normal $x$-$y$ stepping, but with modified electrodes on the outer piezo tube. Characteristics of this angular positioning mode are discussed in 5.

### 2.2 Attachment to field plate

A recent advancement has been the invention of a vacuum-based system for attaching Starbugs to the field plate. In this system, the space between each Starbug's inner and outer piezo tubes is connected via a thin hose to an evacuated chamber (Figure 2). The resultant 'suction' force keeps the Starbug securely attached to the field plate at any angle, while still allowing it to walk. Starbugs do not detach from the plate while positioning because of their small actuator displacements (order 10 µm), which result in negligible airflow. A vacuum pump is required to maintain a sufficiently low pressure within the system. The suction force produced for each Starbug is a function of: i) the area under vacuum; and ii) the pressure difference between a Starbug's interior and the atmosphere. Hence, Starbugs with a larger diameter have a higher suction force, but this force will reduce at altitude due to lower atmospheric pressure.

Vacuum-attached Starbugs have replaced previous designs where a pair of magnets was used to 'clamp' each Starbug to the field plate. This method presented numerous challenges due to the mass of the magnets reducing Starbug performance, particularly at large field plate angles (pointing near horizon). In addition, a low friction transparent coating was required on top of the field plate, which proved difficult to source. The minimum spacing for adjacent Starbugs was also large due to the size of the magnetic assembly. Vacuum Starbugs have dispelled all of these problems.

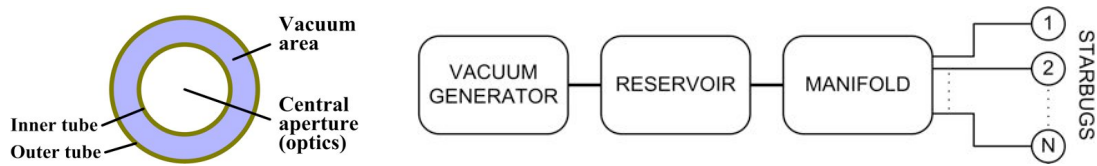

Figure 2: (Top-left) the space between a Starbug's inner and outer tubes is under vacuum, while the inner tube serves as a clear aperture for inserting optics; (top-right) many Starbugs can be connected to the vacuum system via a manifold; (bottom) thin silicone hoses have been used to connect these prototype Starbugs to the vacuum system.

## 3. CONTROL SYSTEM

The current Starbug control system comprises software, drive electronics and a metrology camera for simultaneous closed-loop positioning of multiple Starbugs.

### 3.1 Metrology, software and algorithms

The metrology system tracks the movement of each Starbug using illuminated markers. Current Starbug prototypes have three dedicated metrology fibres for this purpose, arranged in an asymmetric pattern to provide directional as well as positional information (Figure 3). Starbugs are tracked in real-time by the control software, via a camera (or cameras) imaging the entire field. The software uses centroid algorithms and averaging to obtain sub-pixel coordinates for every Starbug's metrology markers and therefore determine the position of the optics at the Starbug's centre. For a telescope instrument, these positions would be measured with reference to fiducial markers on the field plate itself. Camera and telescope distortion models would be included so that Starbugs ultimately move to precise sky coordinates.

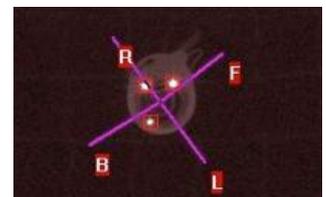

Figure 3: Calibrated movement vectors for each Starbug are recorded relative to three metrology markers.

The control system can characterise the behaviour of each Starbug with a calibration procedure. This may be as simple as driving the Starbug in each of its four directions (forwards, backwards, left, right) for a known number of steps and analysing the resultant movement. The calibration data can be stored in memory and used by the software to predict the Starbug's response during positioning. The algorithms used for routing Starbugs must consider collision avoidance and fibre entanglement issues which, while not a trivial process, will primarily depend on the format of the positioning system. For instruments with a large number of common fibre inputs, one can imagine a uniform distribution of Starbugs across the field plate, each with a defined 'patrol radius' within which it is allowed to move. This reduces the possible collisions between Starbugs and limits the potential for fibre entanglement. In all cases, it is the software system which must keep a record of past moves so that entanglement does not become severe. The simplest case might be that Starbugs have a specified 'home' position, to which they return before a new field configuration is performed.

## 3.2 Electronics

The electronics system (Figure 4) is responsible for providing Starbugs with the high voltage waveforms necessary to drive them in a given direction for a given number of steps, as dictated by the software system. Movement instructions for each Starbug are received and interpreted by a microcontroller, which manages the waveform generation and signal routing for all Starbugs under its control. As described in 2.1, four sinusoidal waveforms are used to produce a Starbug's discrete stepping motion, with its direction of movement controlled by applying the waveforms to different electrodes on the actuators. The four waveforms are generated by a microcontroller and a digital-to-analogue converter (DAC). The DAC output is amplified by power op-amps to produce drive waveforms with amplitudes of up to 200 V, which are then connected to the relevant Starbug electrodes using an array of solid state relays. This means that one set of drive waveforms can be shared by many Starbugs, with their direction controlled by re-routing the signal paths rather than changing the signals themselves. The maximum number of Starbugs that can be driven with one set of waveform amplifiers depends on: i) the size of the Starbugs; ii) the chosen drive voltage and frequency (i.e. step size and speed); and iii) the current rating of the amplifiers. Previous calculations have indicated capacities of order 100 Starbugs.

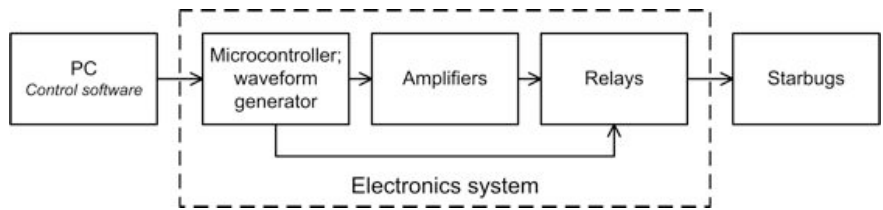

Figure 4: The electronics system provides multiple Starbugs with high voltage drive waveforms.

## 3.3 Scalability

The modular nature of the Starbugs control system means that it is easily scalable. For large numbers of Starbugs, the approach would be to divide these into subgroups, with each subgroup being managed by its own electronics system (Figure 5). The method used for this division is not important; for example, one could group Starbugs based on their payload type, the instrument they feed or the quadrant of the field plate in which they operate. Since only high level commands (Starbug address, direction to move, number of steps to take) are sent by the control software to the electronics system, managing data flow between subgroups would be trivial. Using a grouped approach like this would permit thousands of Starbugs to operate on a single focal surface and would also bolster the redundancy of the system.

A limitation for the maximum number of Starbugs may exist where one wishes to control the entire positioner from a single PC. As the multiplex increases, the software system must maintain centroid coordinates for all Starbugs, in addition to managing collision avoidance and entanglement. Performing this in real-time (or with minimal overheads) could necessitate more than a single desktop computer. A possible solution is to divide the metrology system into subgroups, each managing different areas of the field plate. Indeed, multiple cameras may be needed when imaging large field plates in order to provide a sufficient feedback resolution and frame rate (depending on available technology).

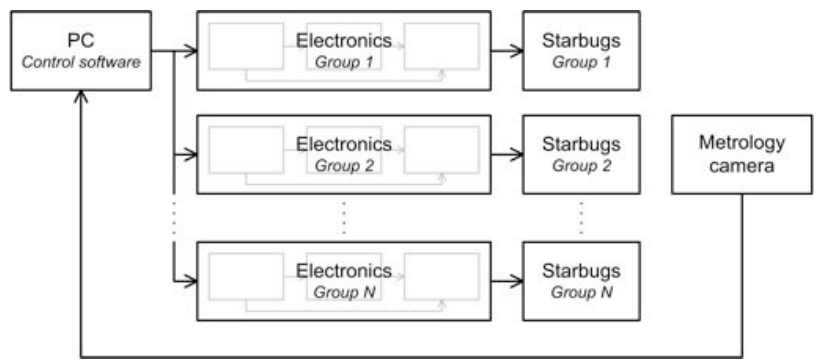

Figure 5: The modular nature of the Starbugs control architecture allows for easy scaling of the system from tens of Starbugs to thousands, with minimal data transfer requirements; multiple metrology cameras may be used for large focal planes.

# 4. LABORATORY TESTING

The measures of Starbug performance presented in this paper are the result of extensive laboratory testing. The AAO has tested various prototype Starbug devices with a purpose-built laboratory demonstrator, described below.

## 4.1 Prototype Starbug designs

Starbugs can be made in various sizes. Their design involves a trade-off of three main parameters: i) maximum speed; ii) maximum load; and iii) minimum spacing (pitch). Three vacuum type Starbug designs have been produced for testing (Figure 6), which will be referred to as 'type-0', 'type-1' and 'type-2' Starbugs for the purpose of this paper. Table 1 shows their specifications. The type-1 Starbug is the current baseline design for the MANIFEST project (see 7).

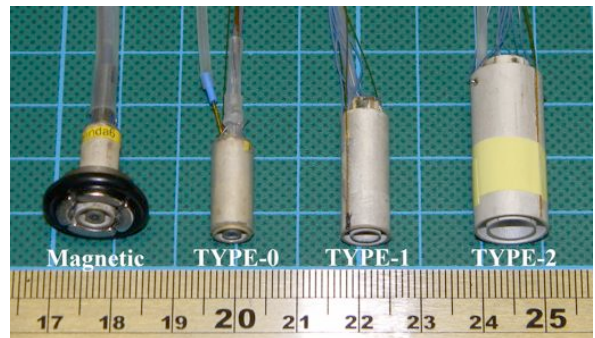

Figure 6: Three vacuum Starbug prototype designs, with the previous magnetic type shown for comparison; scale is in millimetres.

Table 1: Specifications of the three vacuum Starbug designs shown in Figure 6.

|  | Type-0 Starbug | Type-1 Starbug | Type-2 Starbug |
|---|---|---|---|
| Outside diameter | 6.4 mm | 8.0 mm | 12 mm |
| Aperture diameter | 2.2 mm | 4.0 mm | 7.0 mm |
| Height | 20 mm | 25 mm | 30 mm |
| Mass | 2.5 g | 3.9 g | 7.0 g |

## 4.2 Laboratory demonstrator

The AAO has built a dedicated test system for the ongoing development and demonstration of Starbug technology. The Starbugs 'test rig' (Video 2) has a 400 mm removable glass field plate and control electronics for up to 20 Starbugs. A high resolution machine vision camera images the whole field to provide positional feedback to control software running on a PC. The vacuum system uses commercial off-the-shelf components and is programmed to maintain a pressure of ~30 kPa (0.3 atmospheres), to simulate the reduced pressure differential that would occur at high altitude observatories. The entire system is mounted on a motor-driven axle to simulate the pointing of a telescope from zenith to the horizon.

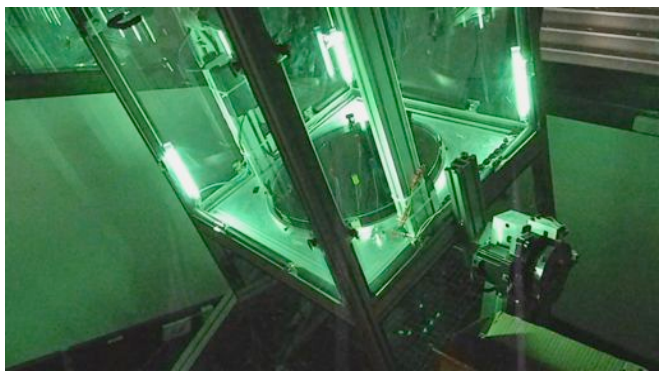

Video 2: The Starbugs 'test rig' simulates several aspects of a real telescope;
http://proceedings.spiedigitallibrary.org/data/Conferences/SPIEP/68021/PSISDG_8450_84501A_ds002.mov

## 5. PERFORMANCE

The results in this section are for the prototype Starbugs described in 4.1. Four key areas of performance are discussed: i) resolution; ii) speed; iii) reconfiguration time; and iv) field plate angle and payload mass.

### 5.1 Resolution

The discrete step size of a Starbug is proportional to the amplitude of the waveform applied to its inner piezo tube (Figure 7), and therefore the minimum step size defines a Starbug's positioning resolution. Testing has shown that step sizes of <4 μm are readily achievable with all Starbug prototypes, in all directions ($\pm x$, $\pm y$). Repeatable sub-micron positioning has also been demonstrated, but was found to be dependent on the quality of assembly. Tests of Starbugs in rotation mode have shown minimum angular step sizes of ~3 arcmins (Figure 8).

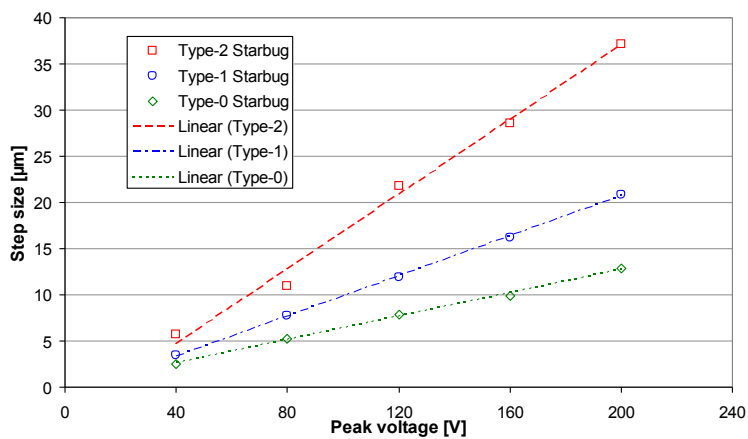

Figure 7: Results show a linear relationship between a Starbug's *x-y* step size and waveform voltage; step sizes of a few microns can be achieved for use in fine positioning; results are for a nominal drive frequency of 100 Hz.

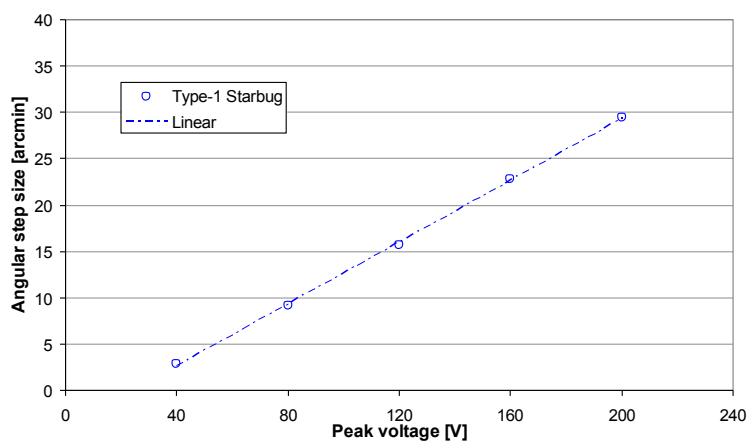

Figure 8: Results show a linear relationship between a Starbug's angular step size and waveform voltage; step sizes of a few arcmins can be achieved for use in fine positioning; results are for a nominal drive frequency of 75 Hz.

For many astronomical applications, it is the metrology system that will limit positioning accuracy. Imaging the entire field with a single camera, for example, will limit resolution with respect to the number of detector pixels and the detector noise. Lens distortion and parallax effects may also introduce errors if not adequately corrected. It follows that for large focal planes it may be necessary to use multiple feedback cameras to achieve the required positioning accuracy.

## 5.2 Speed

For a given step size, how fast a Starbug moves is set by the rate at which it steps, which equals the frequency of the drive waveforms. For example, the results in Figure 7 (above) show a maximum *x-y* step size of ~21 μm for a type-1 Starbug, and so for a waveform frequency of 100 Hz the speed would be 2.1 mm/s (theoretical). Testing has shown that the actual relationship between drive frequency and Starbug speed is dependent on the Starbug design (Figure 9). For most frequencies we see a linear relationship, but at particular frequency values some slippage is apparent, which may be the result of resonant modes within the Starbug. The unpredictable nature of high speed operation for type-1 Starbugs is apparent at 300 Hz. The results show a maximum (stable) speed of 3.7 mm/s for type-1 Starbugs at 200 Hz and 7.5 mm/s for type-2 Starbugs at 300 Hz. Results for Starbug rotation speed are shown in Figure 10. Angular speed increases linearly with frequency for the type-1 prototype that was tested, with a maximum value of 56 deg/s.

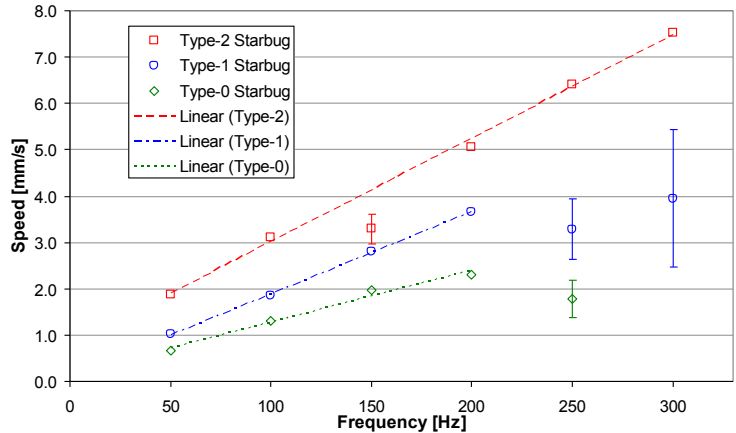

Figure 9: Results show a linear relationship between Starbug speed and lower waveform frequencies; all values were repeatable to <2% except where error bars are shown; results are for a (maximum) drive voltage of 200 V.

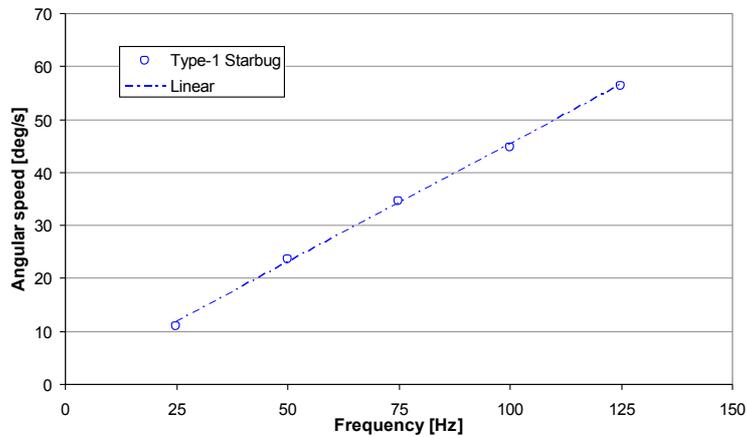

Figure 10: Results show a linear relationship between Starbug angular speed and waveform frequency; behaviour at frequencies higher than 125 Hz was erratic and is not included; results are for a (maximum) drive voltage of 200 V.

High-speed *x-y* movement is particularly desirable when a Starbug must position over long distances, and can be effectively implemented as part of an initial 'coarse' positioning stage. High frequency stepping was found to be less suitable for fine positioning, due to less predictable behaviour over short distances. These findings have naturally led to a two-stage approach to field reconfigurations with Starbugs: the first stage uses a higher frequency and the maximum step size to position a Starbug near its target in the shortest practical time; the second stage then uses a lower frequency and a reduced step size to position the Starbug precisely on its target. The frequencies chosen will depend on the characteristics of the Starbug and the nature of the astronomical instrument (e.g. multiplex, field size).

## 5.3 Reconfiguration time

The time taken to reconfigure a focal plane using Starbugs will depend on more than Starbug performance alone. Collision avoidance and routing algorithms will also influence the total configuration time, as will the distance Starbugs must move and the required accuracy of their final positions. Further, since Starbugs position simultaneously, the time for a complete reconfiguration is naturally dependent on the largest move that must be made. For the case of evenly distributed Starbugs with a defined patrol radius, reconfiguration times can be predicted by scaling experimental data.

Laboratory testing of parallel positioning was performed with 10 type-1 Starbugs under closed loop control. Particular attention was given to positioning performance on a vertical field plate (i.e. telescope pointing at horizon) as, although this angle may not be realised in an instrument, it represents the most challenging conditions for Starbugs in terms of gravitational forces. Figure 11 shows the 'home' positions of the 10 Starbugs and their random target locations for a test reconfiguration (largest Starbug move was ~25 mm). The average time taken to position all Starbugs to within 1/25 pixel (10 µm) of their targets for this case was 42 seconds (at 100 Hz). The reconfiguration can be viewed in Video 3. The test was repeated with different target allocations and all times were consistent to within a few seconds.

The tests described above have proved that simultaneous closed-loop positioning is possible with a simple laboratory control system. From the results of these tests, we can predict the total reconfiguration times for a large scale Starbugs system (Table 2). One of the many merits of Starbug technology is that, for a given Starbug density, multiplying the size of the focal plane will give the same average reconfiguration time. It follows that the average reconfiguration time for an entire field is a function of the patrol radius of the Starbugs.

Table 2: Estimated field reconfiguration times for type-1 Starbugs with different patrol radii, based on results from real reconfigurations in the laboratory; operation at a frequency of 200 Hz is assumed for the estimated figures (based on separate tests of Starbug speed vs. Frequency).

| Starbug patrol radius | Duration of coarse positioning (to ±250 µm) | Duration of fine positioning (to ±10 µm) | Reconfiguration Time (type-1 Starbug) |
|---|---|---|---|
| Measured values (100 Hz coarse and fine positioning) | | | |
| 25 mm | 22 s | 20 s | 42 s |
| Estimated values (200 Hz coarse positioning; 100 Hz fine positioning) | | | |
| 50 mm | 22 s | 20 s | 42 s |
| 100 mm | 44 s | 20 s | 64 s (1.1 mins) |
| 200 mm | 88 s | 20 s | 108 s (1.8 mins) |
| 400 mm | 176 s | 20 s | 196 s (3.3 mins) |

The results in Table 2 were gathered before the creation of a rotation mode for Starbugs. The new rotation mode may allow for more efficient movement over long distances and better implementations of Starbug routing algorithms, yielding faster reconfiguration times for large patrol radii. For example, Starbugs could rotate until their movement axes align with the target position and then execute a single move to get there, rather than taking a 'zig-zig' approach.

There is also the possibility of repositioning some Starbugs while observing, particularly in situations where different groups of objects require different exposure times or are being fed to different instruments. The feasibility of this depends on whether the metrology system (i.e. illuminated markers) would interfere with observations, as open loop Starbug positioning is likely to be unreliable over long distances. Starbugs may also produce unwanted vibrations which could upset spatially sensitive observations. This has not yet been investigated.

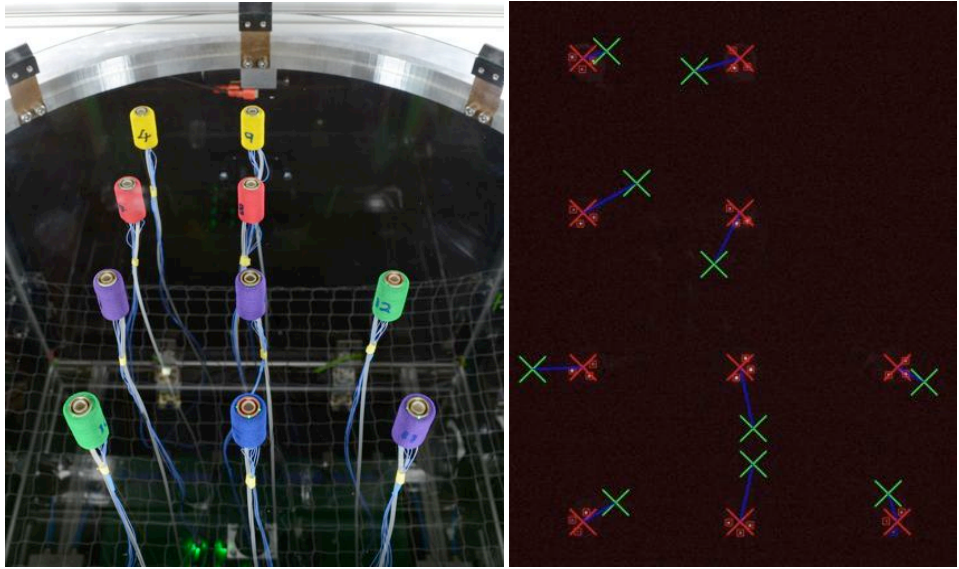

Figure 11: (Left) 10 type-1 Starbugs in their home positions; (right) metrology camera / software view of the field plate showing the Starbug home positions and their target positions (largest move was ~25 mm).

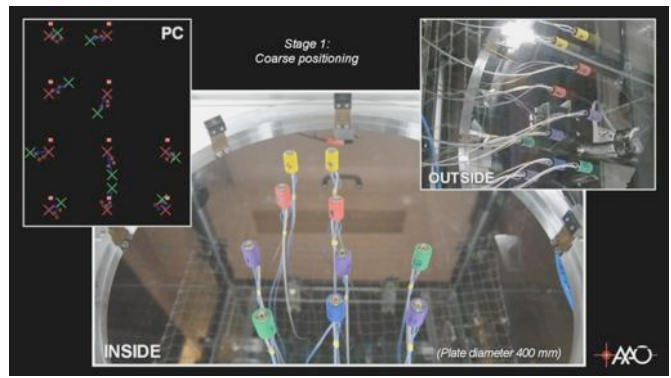

Video 3: Demonstration of a 10 Starbug parallel field reconfiguration on a vertical field plate using type-1 Starbugs; positioning is completed to an accuracy of 10 μm in 42 seconds;
http://proceedings.spiedigitallibrary.org/data/Conferences/SPIEP/68021/PSISDG_8450_84501A_ds003.mov

### 5.4 Field plate angle and payload mass

Various forces will be exerted on a Starbug by its payload as the field plate tilts with telescope pointing. Figure 12 shows that with no additional load (i.e. Starbug moving its own mass and that of its wiring and metrology fibres) there is a measurable decrease in performance as the field plate tilts through 90 degrees. This is due to slippage of the Starbug tubes against the field plate, which will worsen as additional load is applied. For large field plate angles, the dominant mass limit is set by the maximum moment that can be applied before the Starbug 'tips' and detaches from the field plate. Hence, taller and/or narrower Starbugs have a lower limit than shorter and/or wider Starbugs. Wider Starbugs are preferable here as their larger diameter not only tolerates a greater mechanical moment for a given vacuum force, but also provides an increased vacuum force. Of course, this comes at the cost of Starbug pitch.

Theoretical and measured mass limits for different Starbug designs are shown in Table 3, assuming a maximum field plate tilt of 70 degrees from zenith. These results do not consider the stiffness of payloads such as large fibre bundles, which will exert significant additional forces. This highlights a potential restriction with the present state of Starbug technology, that in order to position large and stiff fibre bundles (especially over large distances and at large field plate angles) one needs a Starbug much wider than the bundle itself. The maximum size of a Starbug is not yet known.

Table 3: Measured mass limits for prototype Starbugs on a field plate at 70 degrees from zenith; the values represent the maximum mass in addition to the Starbug's own mass; the bottom row shows the point at which performance is 80% of the no-load value.

|  | **Type-0 Starbug** | **Type-1 Starbug** | **Type-2 Starbug** |
|---|---|---|---|
| Static limit | 14.5 g | 23 g | 67 g |
| Dynamic limit | 2.9 g | 8.5 g | 23 g |
| Limit for 80% performance | **1.5 g** | **4.8 g** | **12 g** |

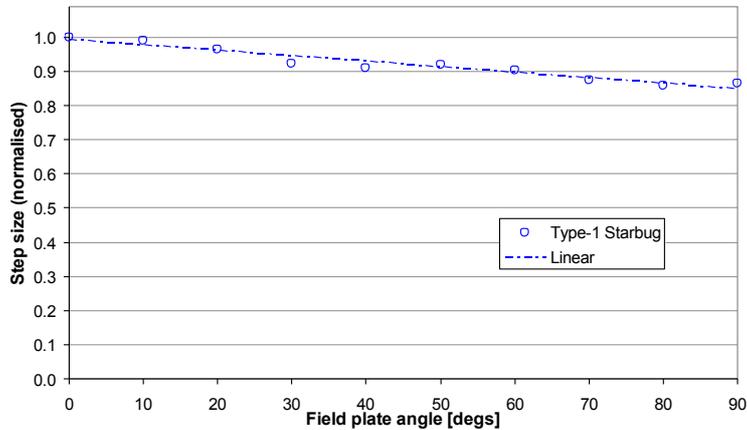

Figure 12: As the field plate angle increases, the step size and hence the speed of the Starbug (moving against gravity) decreases due to slippage; these results are for a type-1 Starbug with no payload other than its own wiring and metrology fibres.

## 6. RELIABILITY AND MAINTENANCE

In retiring the risks of Starbug technology, the AAO has investigated: Starbug lifespan; field plate endurance; and vacuum system robustness. These are discussed below.

### 6.1 Starbug lifespan

Long term 'stress testing' has shown that Starbugs are robust in terms of physical wear and performance over extended periods. Performance data was logged for a single prototype Starbug moving continuously across a glass surface for approximately 48 hours (Figure 13). During this time the Starbug took 10 million discrete steps, divided equally between the $\pm x$ and $\pm y$ directions in successive cycles (moving in a square path); Table 4 shows estimates of the period of use this number of steps may represent for a real multi-object positioner. The results show some initial variation in step size, which may be attributed to heating in the Starbug as a result of the piezo actuators being driven continuously. After this period the performance for all directions is quite stable, varying by ~5% between one cycle and the next in the worst case. These findings suggest not only that Starbugs perform consistently over long periods, but also that they can operate with a duty cycle approaching 100% – much higher than would be required in a real astronomical instrument.

At the time of writing, the same prototype Starbug used in the test above has taken over 30 million steps (see Table 4) and is still performing within 10% of early-life values. No maintenance has been necessary since the Starbug was first assembled, proving that Starbugs can operate for years without the need for replacement. Furthermore, the Starbug control system is immune to small performance variations as long as calibration data is periodically updated.

Table 4: Estimated period of use represented by number of Starbug steps taken for different average move sizes; calculations assume 1000 field reconfigurations per year (8 configurations/night, 125 nights/year).

| **Average move size** | **Usage represented by 10 million steps** | **Usage represented by 30 million steps** |
|---|---|---|
| 50 mm | 4 years | 12 years |
| 100 mm | 2 years | 6 years |
| 200 mm | 1 year | 3 years |
| 400 mm | 6 months | 1.5 years |

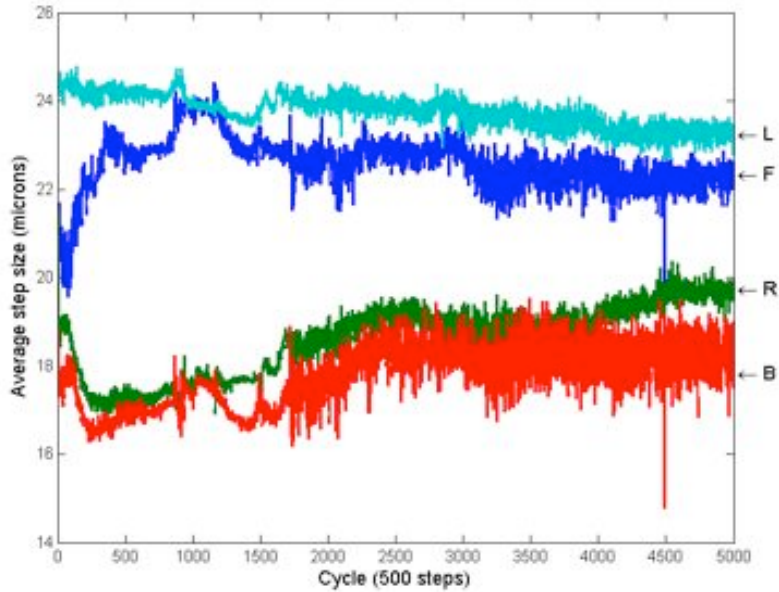

Figure 13: Step size variation of a Starbug under continuous operation for ~48 hrs; each 'cycle' on the x-axis represents 500 steps in all directions (forwards, backwards, left, right), therefore the 5000 cycles shown equals 10 million discrete steps; the results show variation up to cycle ~1700 which may be explained by heating in the actuators (due to continuous operation), after which the short term variation is only a few percent; overall we see no significant long term degradation of performance.

### 6.2 Field plate endurance

Starbugs are predominantly intended to position across a glass field plate, where light passes through the plate before entering the Starbug optics. This means that the field plate must resist mechanical wear as Starbugs walk across its surface, to a degree where the throughput or path of incoming light is not significantly degraded due to wear caused by repeated positioning. Microscope analysis of BK-7 glass samples has confirmed that Starbugs do not visibly damage this material during an equivalent of 5 years' use (at 1000 fields per year). However, testing on a generic anti-reflection coating showed significant wear after this period, and so further research in the area of hard coatings is required.

The wear of the field plate surface and of the Starbugs themselves is dependent on the properties of the materials at their interface; at present, the field plate makes direct contact with a Starbug's piezoceramic actuators. Future work will look at the potential advantages of adding other materials to the actuator tips, either as plating or as Starbug 'shoes'.

### 6.3 Vacuum system robustness

Starbugs require an active vacuum system in order to remain attached to the field plate. It is therefore critical that this vacuum system is robust, reliable and fault-tolerant. To this end, we envisage a design featuring inherent redundancy and methods for fault isolation. An advantageous characteristic of Starbugs is that their micro-stepping motion produces negligible leakage during operation. This is because only a small gap (order 10 μm) is present between the field plate and the Starbug while it is stepping. When a Starbug is stationary, its polished surface provides an effective seal against the field plate and almost no leakage is present. It has been demonstrated that a single Starbug will remain attached to a field plate for over one hour when connected to a small vacuum reservoir, and so with adequate capacity a large Starbugs system could tolerate complete power failures. In addition, the vacuum in each Starbug is fed by a narrow tube which naturally restricts airflow, ensuring that multiple Starbugs do not lose suction in the event that one detaches from the plate. Starbugs have also been shown to remain attached to a horizontal field plate during simulated earthquakes with accelerations of up to 3g; the peak acceleration for a 2%, 500 year earthquake event at the GMT instrument platform is stated as 1.38g.

# 7. STARBUGS FOR MANIFEST

The AAO is planning to build MANIFEST, a multi-object fibre positioning facility instrument for the 24.5 m Giant Magellan Telescope (GMT)[7], which can feed three proposed instruments[8]: the wide-field multi-object spectrograph GMACS; the high resolution spectrograph G-CLEF; and the near infrared multi-object spectrograph NIRMOS. The positioning requirements for these instruments across the GMT's Gregorian focal plane make Starbugs the AAO's technology of choice for MANIFEST. Much of the recent progress with Starbugs has been the result of a feasibility study and intensive R&D program for MANIFEST, funded by the Giant Magellan Telescope Organization (GMTO).

The mechanical design of MANIFEST (Figure 14) has a ~1.3 m diameter glass field plate matching the focal plane's 3275 mm spherical radius of curvature. Underneath this field plate are several hundred Starbugs, attached by a vacuum and evenly distributed across the field and with a defined 'patrol radius'. The vacuum feed, electrical wiring and back-illuminated metrology fibres are coupled to each Starbug via a connector board underneath the field plate. The system positions various payloads including single fibres, small fibre bundles (e.g. 19 fibres) and large Integral Field Units (IFUs). It is therefore likely that a range of Starbug designs will be required. Average field reconfiguration times for MANIFEST are expected to be three minutes or less. First light for the GMT is projected to be in 2020.

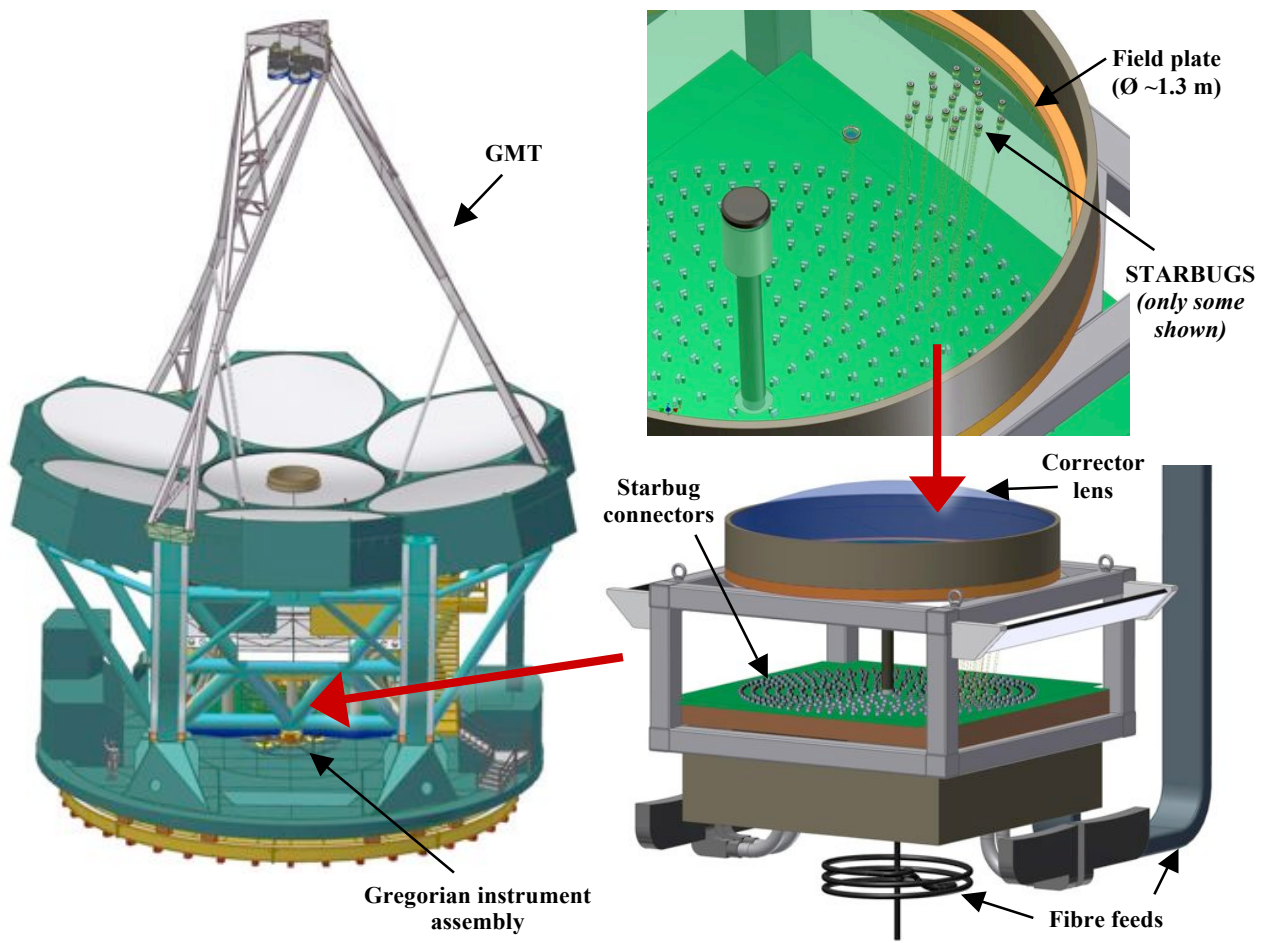

Figure 14: 3D models of: (top-right) Starbug devices on the MANIFEST field plate; (bottom-right) the whole of the MANIFEST instrument; (left) the Giant Magellan Telescope

# 8. CONCLUSIONS

Extensive R&D has transported Starbugs from the realm of a conceptual technology to that of a proven one. Fast and precise multi-object positioning with Starbugs has been successfully demonstrated in a simulated telescope environment and has proved reliable and repeatable. Furthermore, the Starbug itself has emerged as a robust device which can perform many thousands of reconfigurations without replacement or maintenance. In addition to this, the capabilities of the technology have been boosted by the use of vacuum-driven field plate attachment and the invention of a rotation mode for angular as well as orthogonal $x$-$y$ positioning.

Further research will foster a better understanding of a Starbug's dimensional limits for large and/or heavy payloads and also for small pitch, high density applications with lightweight payloads. Another important area of study will be in optimising the mechanical interface of Starbugs with the field plate through the use of 'shoes' or plating techniques on their piezoceramic actuators. The AAO is planning to demonstrate Starbugs on the sky with a prototype positioning system, which will aim to configure a range of fibre payloads and pave the way to designing Starbugs for MANIFEST. Starbugs are also candidates for the AAO's multi-IFU instrument HECTOR[9], although further work is necessary to properly test the compatibility of Starbugs with HECTOR's large and rigid hexabundle assemblies. Outside the requirements of existing instrument proposals, there is scope to adapt Starbugs for use with other focal plane formatting techniques such as deployable pick-off mirrors or sub-field imaging. In all of these cases, it is the Starbug's marriage of simplicity and versatility that makes it a truly attractive option for so many applications.